\documentclass[aps,prl,superscriptaddress,reprint,showpacs,amssymb,amsmath,floatfix,citeautoscript,longbibliography]{revtex4-1}

\usepackage{color}       
\usepackage{graphicx}                             
\usepackage{epstopdf}
\usepackage{color}
\definecolor{red}{rgb}{1.00,0.00,0.00}

\date{\today}

\begin{document}

\title{Absence of magnetic-proximity effect at the interface of Bi$_2$Se$_3$ and (Bi,Sb)$_2$Te$_3$ with EuS}

\author{A. I. Figueroa}
\email{adriana.figueroa@icn2.cat}
\affiliation{Catalan Institute of Nanoscience and Nanotechnology (ICN2), CSIC and BIST, Campus UAB, Bellaterra, 08193 Barcelona, Spain}

\author{F. Bonell}
\affiliation{Catalan Institute of Nanoscience and Nanotechnology (ICN2), CSIC and BIST, Campus UAB, Bellaterra, 08193 Barcelona, Spain}

\author{M. G. Cuxart}
\affiliation{Catalan Institute of Nanoscience and Nanotechnology (ICN2), CSIC and BIST, Campus UAB, Bellaterra, 08193 Barcelona, Spain}
\affiliation{Universitat Aut\`{o}noma de Barcelona (UAB), Bellaterra 08193, Spain}

\author{M. Valvidares}
\affiliation{ALBA Synchrotron Light Facility, Barcelona 08290, Spain}

\author{P. Gargiani}
\affiliation{ALBA Synchrotron Light Facility, Barcelona 08290,
Spain}

\author{G. \surname{van der Laan}}
\affiliation{Diamond Light Source, Harwell Science and Innovation
Campus, Didcot, OX11~0DE, United Kingdom}

\author{A. Mugarza}
\affiliation{Catalan Institute of Nanoscience and Nanotechnology (ICN2), CSIC and BIST, Campus UAB, Bellaterra, 08193 Barcelona, Spain}
\affiliation{Instituci\'{o} Catalana de Recerca i Estudis Avan\c{c}ats (ICREA), Barcelona 08010, Spain}

\author{S. O. Valenzuela}
\email{SOV@icrea.cat}
\affiliation{Catalan Institute of Nanoscience and Nanotechnology (ICN2), CSIC and BIST, Campus UAB, Bellaterra, 08193 Barcelona, Spain}
\affiliation{Instituci\'{o} Catalana de Recerca i Estudis Avan\c{c}ats (ICREA), Barcelona 08010, Spain}

\begin{abstract}
%
We performed x-ray magnetic circular dichroism (XMCD) measurements on heterostructures comprising topological insulators (TIs) of the (Bi,Sb)$_2$(Se,Te)$_3$ family and the magnetic insulator EuS. XMCD measurements allow us to investigate element-selective magnetic proximity effects at the very TI/EuS interface. A systematic analysis reveals that there is neither significant induced magnetism within the TI nor an enhancement of the Eu magnetic moment at such interface. The induced magnetic moments in Bi, Sb, Te, and Se sites are lower than the estimated detection limit of the XMCD measurements of $\sim\!10^{-3}$ $\mu_\mathrm{B}$/at.
%
\end{abstract}

\date{\today}
\maketitle

The observation of the quantum anomalous Hall effect (QAHE) in magnetically-doped topological insulators (TI) has raised significant interest for applications in metrology \cite{Chang2013,Chang2015}.
Test measurements demonstrating QAHE quantization accuracy at a
record level of better than 1 part-per-million have recently been
reported \cite{Gotz2018}. However, the magnetic impurities are
incorporated  randomly in the TI lattice and the resulting disorder
presumably reduces the temperature and currents at which the QAHE
fully develops. It also leads to the observation of
superparamagnetic phases and multidomain magnetization switching,
which coincide with the QAHE and are not well understood
\cite{Winnerlein2017}. In this context, imprinting magnetism by
proximity effects in thin film heterostructures appears as a
promising alternative
\cite{LiPRL2015,JiangNanoLett2015,KatmisNat2016, TangSciAdv2017}
that could ensure the preservation of the TI crystalline quality and
its bulk insulating properties. Magnetic proximity effects in
heterostructures comprising TIs and magnetic insulators (MI) have
been intensively investigated using optical methods, electronic
transport measurements, polarized neutron reflectivity, and muon
spin spectroscopy \cite{WeiPRL2013, KatmisNat2016, LeeNatComms2016,
JiangNanoLett2015, LiPRB2017, TangSciAdv2017, Krieger_PRB2019,
MogiPRL_2019, Watanabe_2019, Mathimalar_2019, Pereira_PRM2020}.
Studied heterostructures include Bi$_{2}$Se$_3$/EuS,
(Bi,Sb)$_{2}$Te$_3$/EuS,
Y$_3$Fe$_5$O$_{12}$/(Bi$_x$Sb$_{1-x}$)$_2$Te$_3$,
Tm$_3$Fe$_5$O$_{12}$/(Bi$_x$Sb$_{1-x}$)$_2$Te$_3$,
Fe$_3$O$_4$/Bi$_2$Te$_3$, Cr$_2$Ge$_2$Te$_6$/(Bi,Sb)$_2$Te$_3$, and
(Zn,Cr)Te/(Bi,Sb)$_2$Te$_3$/(Zn,Cr)Te, with the QAHE only reported
in the latter \cite{Watanabe_2019}. However, the most intriguing
results have been arguably observed in Bi$_{2}$Se$_3$/EuS, where
experiments using spin-polarized neutron reflectivity
\cite{KatmisNat2016} suggest an enhanced magnetic signal within
Bi$_2$Se$_3$. They also indicate the persistence of large
interfacial ferromagnetism that extends 2 nm into the TI interface
up to room temperature, even though isolated EuS orders
ferromagnetically only below $17$ K \cite{KatmisNat2016}. Such
results are remarkable and would make Bi$_2$Se$_3$/EuS a leading
candidate for the observation of topological magneto-electric
phenomena and QAHE at high temperatures.

However, while the number of articles reporting
signatures of ferromagnetic behavior in TI/EuS interfaces keeps on
growing, topological phenomena have yet to be observed. QAHE has
been systematically demonstrated in Cr- and V-doped
(Bi,Sb)$_2$Te$_3$ \cite{Chang2013,Chang2015} despite that the
ferromagnetic state appears to be much weaker than in
Bi$_2$Se$_3$/EuS. In addition, first-principles calculations of the
electronic band structure and  magnetic ordering of the
Bi$_2$Se$_3$/EuS interface did not find an induced magnetic moment
in the TI or even a significant enhancement of the local magnetic
moment of Eu \cite{EremeevJMMM2015, KimPRL2017}. The discrepancy
between experiments and  theoretical analysis in combination with
the absence of direct experimental proof of locally induced
magnetism in the TI underscores the need of further experimental
investigations.

Electronic and magnetic information regarding microscopic interactions can be obtained using x-ray absorption spectroscopy (XAS) and x-ray magnetic circular dichroism (XMCD). These techniques have been successfully implemented to study magnetically-doped TIs \cite{figueroaPRB2014, LiuACSNano2015, HarrisonJPhys2015, HarrisonAPL2015,Liu_NanoLett2015, baker2015magnetic, Ye2015, DuffyPRB2017, Russmann_2018} and---most importantly---are able to detect element-specific magnetic moments that are induced in the host TI lattice \cite{Ye2015, Russmann_2018}. In this Letter, we report XAS and XMCD measurements on TI/EuS interfaces, with TIs of the (Bi,Sb)$_2$(Se,Te)$_3$ family. In contrast to previous investigations, our systematic study provides fundamental local and element-selective information of the magnetic moments associated to Eu, Bi, Sb, Se, and Te atoms. In agreement with theoretical reports, we find no indication of proximity-induced magnetism in the TI. This suggests that the magnetic signatures in TI/MI heterostructures which have been reported do not originate from induced ferromagnetic order in the TI atoms at the TI/MI interface. 

The TI/EuS heterostructures were grown on single-crystal BaF$_2$(111) substrates by molecular beam epitaxy under ultrahigh-vacuum (UHV) with a base pressure $\sim\!1\times10^{-10}$ Torr. Bi$_{2}$Se$_3$, Bi$_{2}$Te$_3$, Sb$_{2}$Te$_3$, and (Bi,Sb)$_{2}$Te$_3$ thin films with a thickness of 10 nm were grown by co-evaporation of elemental Bi, Sb, Se, and Te (6N purity), following the procedure in Ref. \onlinecite{Bonell2017}. After growing the TI layer, the samples were transferred in UHV to a second chamber where a layer of 5-nm thick EuS was deposited at room temperature using an electron-beam evaporator, as in Ref. \onlinecite{KatmisNat2016}. A protective 2-nm thick Al capping layer was subsequently deposited in the same chamber and allowed to oxidize in air. Additional Sb$_{2}$Te$_3$/EuS, Sb$_{2}$Te$_3$/Bi$_{2}$Se$_3$/EuS, and Bi$_{2}$Se$_3$/EuS(\textit{x}) with \textit{x} = 1 and 5 nm samples were capped with a 20-nm thick Se layer grown \textit{in-situ}, which was desorbed in the UHV preparation chamber for XAS measurements by heating the sample to $\sim\!180^{\circ}$C. The structural and crystal quality of the films was assessed using \emph{in-situ} reflection high energy electron diffraction and \emph{ex-situ} x-ray diffraction and x-ray photoelectron spectroscopy (XPS) revealing epitaxial growth and sharp interfaces \cite{SupplInfo}. The macroscopic magnetic properties were studied by SQUID magnetometry \cite{SupplInfo}.

XAS was performed using a 6-T magnet on beamline BL-29 (BOREAS) at
the ALBA synchrotron (Spain), which provides a UHV sample
environment with a base temperature of $\sim$3~K. Measurements were
focused on the $M_{4,5}$ edges of Eu (1110-1170~eV), Te
(570-590~eV), Bi (2560-2700~eV), and Sb (525-545~eV), and $L_{2,3}$ edges of Se (1420-1460~eV). The XMCD signal was obtained
by subtracting XAS spectra with the photon helicity vector
antiparallel and parallel to the applied magnetic field.
Measurements used total-electron-yield (TEY) detection, where the
drain current was taken from the sample to the ground. The
magnetic field was applied along the x-ray beam for two different
geometries: normal incidence (perpendicular to the sample plane) and
grazing incidence (30$^\circ$ off the sample plane).

\begin{figure}[tb!]
\begin{center}
\centerline{
\includegraphics[width = 8cm]{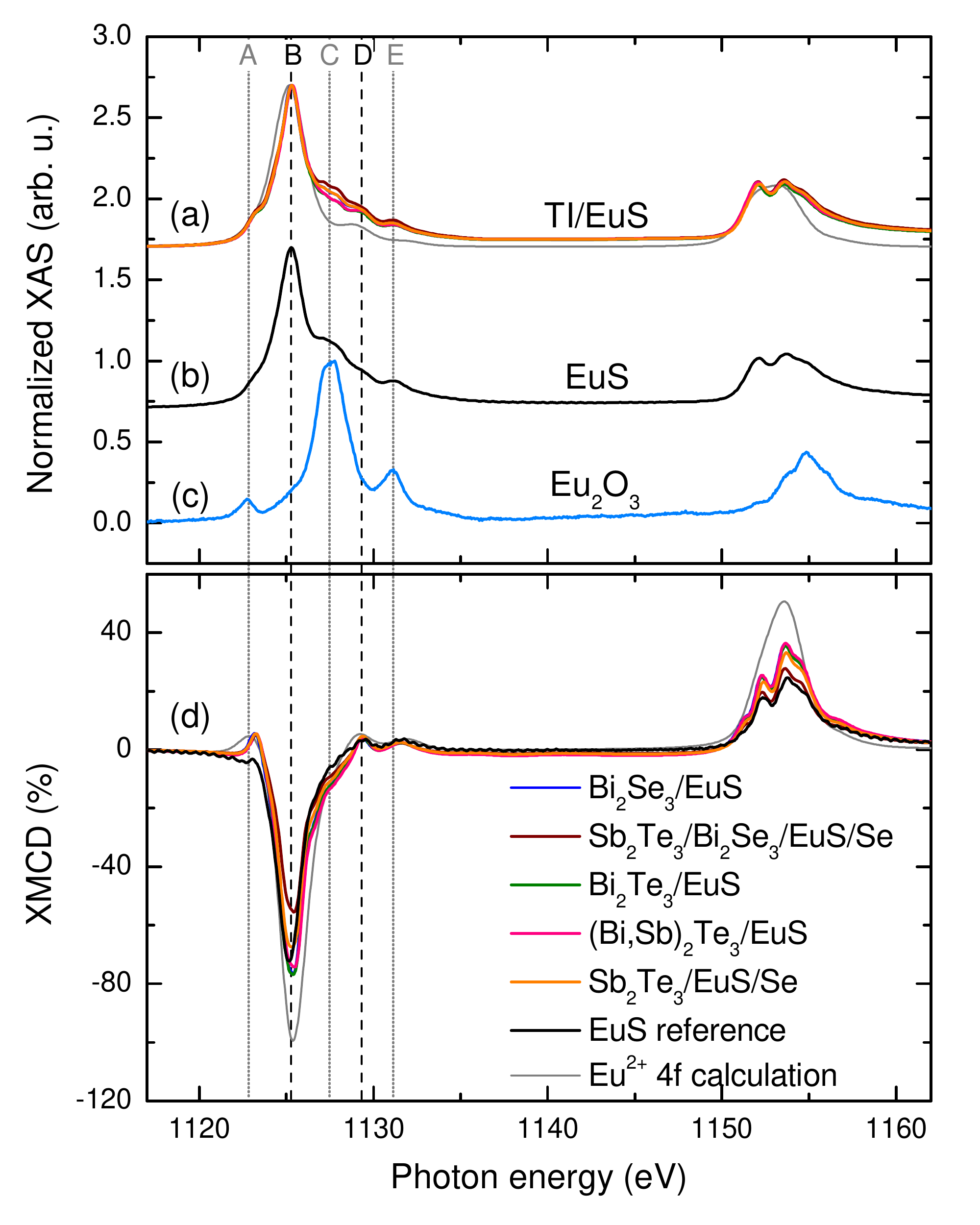}
}
 \caption{(a)
Averaged XAS spectra at the Eu $M_{4,5}$ edges for Bi$_2$Se$_3$/EuS,
Bi$_2$Te$_3$/EuS, (Bi,Sb)$_2$Te$_3$/EuS, Sb$_2$Te$_3$/EuS,
Sb$_2$Te$_3$/EuS/Se samples and comparison with the spectrum
calculated for Eu$^{2+}$ $4f^7$, as well as (b) EuS and (c)
Eu$_2$O$_3$ references. Reference spectra have been vertically
shifted for clarity. (d) XMCD spectra for all samples and
references. Dashed vertical lines mark the main features for Eu$^{2+}$ (peaks
B and D) and Eu$^{3+}$ (peaks A, C, and E) around the $M_5$ edge.
Measurements were performed at 3 K under 2 T magnetic field applied
in grazing incidence.} \label{Fig_EuXMCD}
\end{center}
\end{figure}

Figure \ref{Fig_EuXMCD} depicts XAS and XMCD spectra at the Eu
$M_{4,5}$ edges for grazing incidence at 3 K under 2 T magnetic
field. The $M_5$ edge peak maximum intensity is normalized to unity
in order to compare the spectra from different samples. XMCD is
expressed in percentage of the average XAS. The lineshape and
amplitude of the normalized XAS and XMCD signals for all TI/EuS
samples [Fig.~\ref{Fig_EuXMCD}(a) and (d)] are nearly identical,
which indicates the same electronic and magnetic state of the Eu
atoms. The Eu $M_{4,5}$ edges correspond to $3d\rightarrow4f$
electronic transitions. In order to identify the Eu electronic
state, the results are compared with the calculated spectra for
Eu$^{2+}$ $4f^7$ (Fig.~\ref{Fig_EuXMCD}(a) and Supplementary
Material \cite{SupplInfo}) and the measured spectra in two reference
samples: a 5-nm thick EuS thin film and Eu$_2$O$_3$ in powder form
[Fig.~\ref{Fig_EuXMCD}(b) and (c), respectively]. The comparison
reveals that spectra for all TI/EuS samples display the main
features (peaks B and D) for Eu$^{2+}$, which is consistent with the
expected electronic state for ferromagnetic EuS. Weak features
corresponding to Eu$^{3+}$ (peaks A, C, and E) are also visible,
revealing some degree of Eu oxidation (Eu$_2$O$_3$ phase) at the
surface of the sample, which is also observed and estimated to be
between 0.35-0.47 nm thick by XPS \cite{SupplInfo}. Eu$_2$O$_3$ is
not ferromagnetic, therefore it does not contribute to the XMCD
signal \cite{NegusseJAP2009}.

XMCD is a powerful tool that yields the spin and orbital moments of
the probed elements through the application of the magneto-optical
sum rules \cite{TholePRL1992, CarraPRL1993}. For the rare earths,
however, the sum rule analysis is less straightforward
\cite{vanderlaan1999, van2014x}. Since the XMCD intensity scales
with the magnetization \cite{van2014x}, values for the Eu $4f$
magnetic moments can be directly obtained by comparing the $M_5$
XMCD asymmetry with the expected theoretical value. This asymmetry
$A$ is the maximum value of (XMCD)/(XAS sum), where XMCD and XAS are
the difference and the sum, respectively, of the right- and left-
circularly polarized XAS spectra at the Eu $M_5$ edge.

Table \ref{tab:Eu} lists the magnetic moments derived from $A$ for
each sample at 3 K ($\sim$10~K for
Bi$_{2}$Se$_3$/EuS(\textit{x}) with \textit{x} = 1 and 5 nm
samples). The calculated XMCD spectrum in Fig.~\ref{Fig_EuXMCD}(d)
yields the theoretical asymmetry $A_\mathrm{theory} = -0.497$, which
corresponds to the Hund's rule ground state (GS) value for Eu$^{2+}$
$4f^7$ ($S$ = 7/2, $L$ = 0, $J$ = 7/2), with an effective magnetic
moment $\mu_{\mathrm{eff}}^{\mathrm{GS}} =$ 7.93 $\mu_\mathrm{B}$.
Because $L=0$, this is a pure spin moment. Table \ref{tab:Eu} shows
a reduction of $\mu_\mathrm{eff}$ of about 30-35\% for all samples
with respect to $\mu_{\mathrm{eff}}^{\mathrm{GS}}$, which could be
due to the presence of a small crystal field acting on the Eu $4f$
electrons, as observed for other rare earths
\cite{FigueroaJMMM2017}. These values are well in agreement with the
magnetic moments of Eu in Bi$_2$Se$_3$/EuS calculated by density
functional theory \cite{KimPRL2017}, also included in Table
\ref{tab:Eu}. No significant enhancement in the Eu $4f$ magnetic
moments of our TI/EuS samples is found compared to that of the
isolated EuS layer. This observation is consistent with theoretical
reports \cite{KimPRL2017,EremeevJMMM2015}, but contrary to
experimental findings using polarized neutron reflectometry
\cite{KatmisNat2016}.

\begin{table}
\caption{\label{tab:Eu} Effective magnetic moments for the Eu$^{2+}$
$4f^7$ state in TI/EuS samples as derived from the Eu $M_5$ XMCD
asymmetry. Measurements were recorded at 3 K, except
for those marked with *, which were recorded at $\sim$10~K. The
local spin magnetic moment of Eu in Bi$_2$Se$_3$/EuS for the atomic
ground state and as calculated by DFT \cite{KimPRL2017} are included
for comparison.}
\begin{ruledtabular}
\begin{tabular}{lc}
Sample&$\mu_\mathrm{eff}$($\mu_\mathrm{B}$/atom)\\
\hline
EuS & 5.8 $\pm$ 0.3 \\
Sb$_{2}$Te$_3$/EuS/Se & 5.4 $\pm$ 0.3 \\
Bi$_{2}$Te$_3$/EuS & 6.1 $\pm$ 0.4 \\
(Bi,Sb)$_{2}$Te$_3$/EuS & 5.9 $\pm$ 0.3 \\
Bi$_{2}$Se$_3$/EuS & 6.1 $\pm$ 0.4 \\
Sb$_{2}$Te$_3$/Bi$_{2}$Se$_3$/EuS/Se & 4.4 $\pm$ 0.4 \\
Bi$_{2}$Se$_3$/EuS (1 nm)/Se* & 4.8 $\pm$ 0.4 \\
Bi$_{2}$Se$_3$/EuS (5 nm)/Se* & 5.3 $\pm$ 0.4 \\
Atomic Hund's rule ground state & 7.93 \\
DFT calculation \cite{KimPRL2017} & 6.94\\
\end{tabular}
\end{ruledtabular}
\end{table}

\begin{figure}[tb!]
\begin{center}
\centerline{
\includegraphics[width = 9 cm]{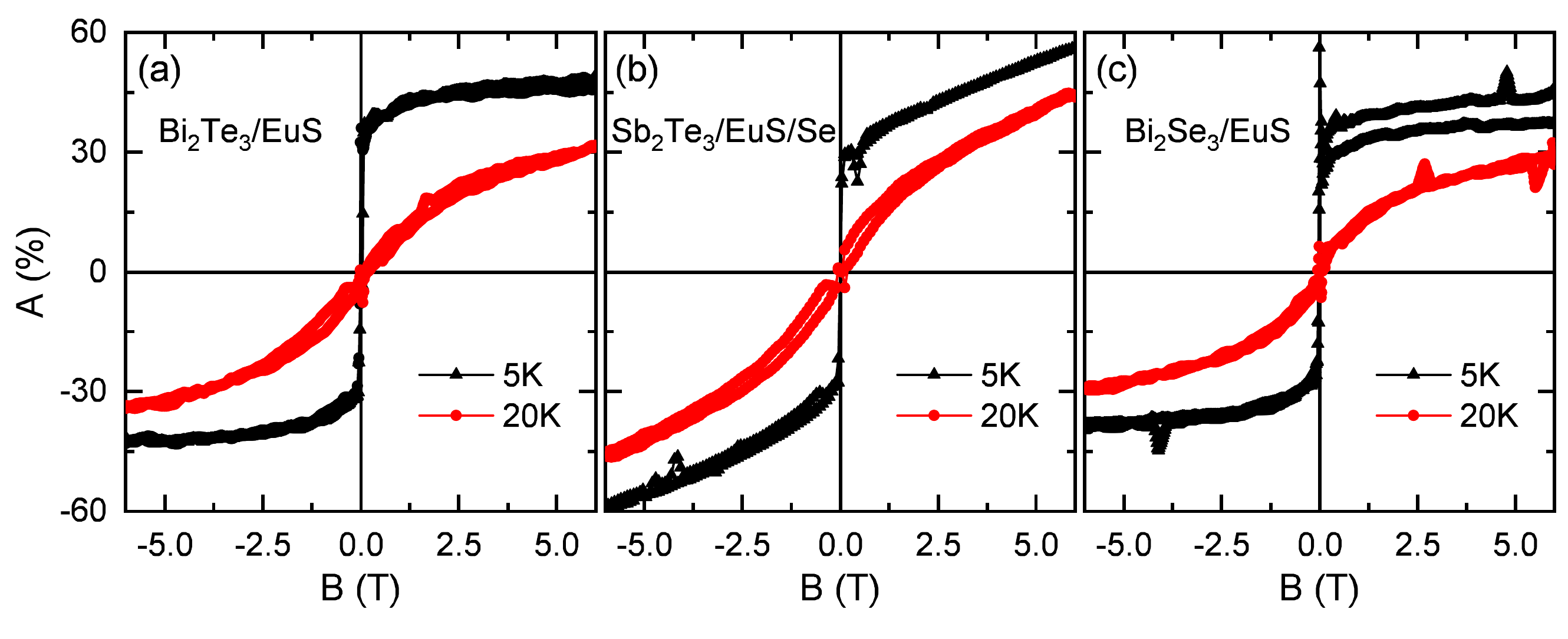}
} \caption{Hysteresis loops recorded at the Eu $M_5$ edge for (a)
Bi$_{2}$Te$_{3}$/EuS, (b) Sb$_{2}$Te$_{3}$/EuS/Se, and (c)
Bi$_{2}$Se$_{3}$/EuS samples, measured in grazing incidence at 3 K
(black triangles) and 20 K (red circles).} \label{Fig_Hyst}
\end{center}
\end{figure}

Magnetic hysteresis loops were recorded by fixing the energy at the
maximum of the Eu $M_5$ XMCD signal (1125.4 eV) while sweeping the
field. Figure \ref{Fig_Hyst} shows loops measured for
Bi$_2$Te$_3$/EuS, Sb$_2$Te$_3$/EuS/Se and Bi$_2$Se$_3$/EuS in
grazing incidence. Values of $A$ at saturation are very similar in
all cases. This is consistent with Eu atoms having the same
electronic and magnetic state regardless of the TI they are in
contact with. The square shape of these hysteresis loops
demonstrates that Eu $4f$ states are ferromagnetic at low
temperatures (3 K) but they become paramagnetic at 20 K. This is in
agreement with results of SQUID magnetometry, which indicate
ferromagnetic ordering below $T_\mathrm{C}\approx 15$ K
\cite{SupplInfo}. Therefore, we do not observe signs of
room-temperature ferromagnetic behavior in these TI/EuS systems, in
contrast to previous reports \cite{KatmisNat2016}.

The magnetic anisotropy of the Eu layer in the TI/EuS system was analyzed by comparing the XMCD spectra recorded in normal and grazing geometries at 2 T and at remanence (see Fig.~S5 in Supplemental Material \cite{SupplInfo}). At 2 T, curves for normal
and grazing geometries overlap for all TI/EuS samples. Without magnetic field, we observe remanence only in samples with the TI underlayer; the isolated EuS layer shows no XMCD signal at zero field. This behavior can be ascribed to the large spin-orbit interaction that is intrinsic to the TI \cite{KimPRL2017}. Moreover, there is a significantly larger remanence in the grazing geometry for all TI/EuS samples, which is indicative of a favorable magnetization axis in the plane of the sample. In-plane anisotropy is also observed macroscopically by SQUID magnetometry (see Fig.~S4 in Supplemental Material \cite{SupplInfo}). Previous
experiments on Bi$_{2}$Se$_{3}$/EuS demonstrated perpendicular anisotropy in EuS \cite{KatmisNat2016}. Theoretical calculations have shown that topological surface states and their large intrinsic spin-orbit coupling may explain this observation \cite{KimPRL2017}.
However, the orientation of the magnetization depends strongly on the strain at the interface. When the EuS lattice is relaxed, as is the case for our TI/EuS films (see Fig.~S1 in Supplemental Material \cite{SupplInfo}), it is expected that the magnetization remains in plane \cite{KimPRL2017}.

\begin{figure}[tb!]
\begin{center}
\centerline{
\includegraphics[width = 8 cm]{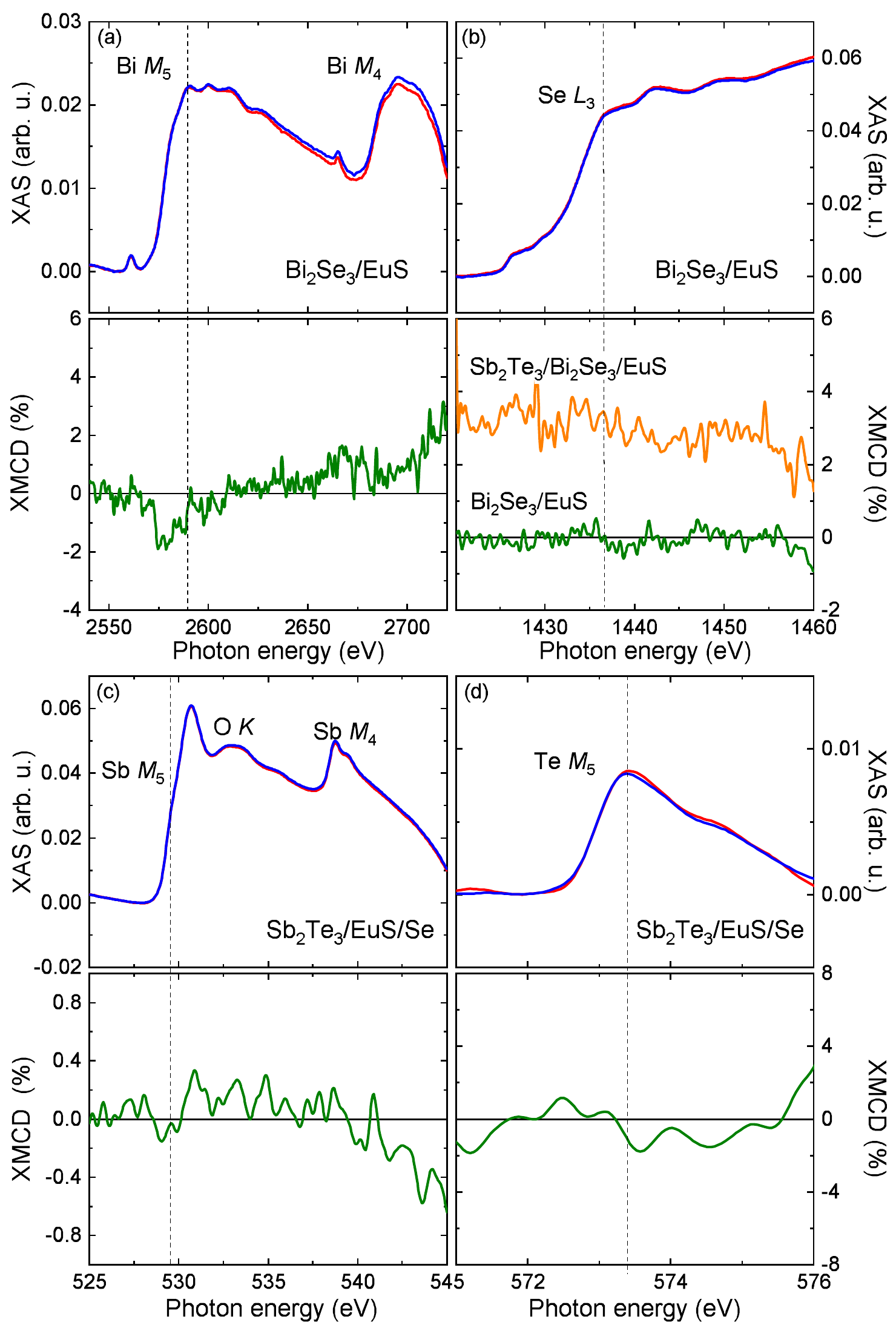}
}
\caption{Experimental spectra for left- and right-circularly polarized x-rays. XAS (top panel) and XMCD (bottom panel) at the (a) Bi $M_{4,5}$ and (b) Se $L_{3}$ edges for Bi$_{2}$Se$_{3}$/EuS and the (c) Sb $M_{4,5}$ and (d) Te $M_{5}$ for Sb$_{2}$Te$_{3}$/EuS. The XMCD signal at the Se $L_{3}$ for the Sb$_{2}$Te$_{3}$/Bi$_{2}$Se$_{3}$/EuS sample has been included in (b) for comparison (vertically shifted for clarity). Measurements were performed at 3 K under a magnetic field of 2 T applied along the normal. Dashed lines mark the energy at which the XMCD features are expected to appear.} \label{Fig_NonMagAt}
\end{center}
\end{figure}

Next we will focus on the magnetic proximity effects. XMCD measurements at
accessible absorption edges of Bi, Te, Sb, and Se elements were
performed in search of induced magnetism on these non-magnetic
atoms. Figure \ref{Fig_NonMagAt} depicts XAS and XMCD spectra at the
Bi $M_{4,5}$ and Se $L_{3}$ edges for Bi$_{2}$Se$_{3}$/EuS and at
the Sb and Te $M_{5}$ edges for Sb$_{2}$Te$_{3}$/EuS. No XMCD signal
is apparent in any of these absorption edges probed, remaining
within the noise level.

TEY-detected XAS for soft x rays is a surface sensitive technique
with the majority of electrons originating from within 3-5 nm
\cite{StohrBook2006, van2014x}. This implies that the XAS signal not
only arises from the uppermost region of the TI, which is in contact
with EuS, but also from the bulk of the film with a contribution
that decreases exponentially with depth. In order to verify the
absence of proximity-induced magnetism, we performed XMCD
measurements at the Se $L_{3}$ and Bi $M_{4,5}$ edges in a
Sb$_{2}$Te$_{3}$/Bi$_{2}$Se$_{3}$/EuS heterostructure. The
Bi$_{2}$Se$_{3}$ layer is 1 nm thick, ensuring that the recorded XAS
signal originates from the very Bi$_{2}$Se$_{3}$/EuS interface. XMCD
at the Se $L_{3}$ for this sample is plotted in
Fig.~\ref{Fig_NonMagAt}(b) (the Bi signal on this sample, not shown,
is very weak). In agreement with the results on the
Bi$_{2}$Se$_{3}$/EuS, no XMCD signal is found at this absorption
edge.

Non-negligible Te and Sb XMCD signals have been reported for thin
films of Cr-doped Sb$_2$Te$_3$ \cite{DuffyPRB2017, IslamPRB2018} and
bulk Cr-, V-, and Mn-doped Sb$_2$Te$_3$ \cite{DuffyPRB2017,
IslamPRB2018} and Cr-doped (Bi,Sb)$_2$Te$_3$ \cite{Ye2015,
YePRB2019}. In all these cases, XMCD data was recorded in TEY for
\textit{in-situ} cleaved samples. TEY detected XAS and XMCD signal
for our TI/EuS heterostructures have poor statistics compared to
those reported for \textit{in-situ} cleaved samples due to the 5 nm
EuS and 2 nm Al on top of the TI. Despite the noise level of the
data, application of the magneto-optical sum rules
\cite{TholePRL1992, CarraPRL1993} to the XAS and XMCD signal in
Fig.~\ref{Fig_NonMagAt} can provide an upper limit for the induced
magnetic moment on Bi, Sb, Se, and Te atoms in contact with EuS
\cite{SupplInfo}. We found values of the spin moment, $S_z\approx
10^{-3}$ $\mu_\mathrm{B}$ per Bi and Se atom in the
Bi$_{2}$Se$_{3}$/EuS samples and $\sim\!10^{-3}$ $\mu_\mathrm{B}$ per
Sb and $\sim\!10^{-4}$ $\mu_\mathrm{B}$ per Te in the
Sb$_2$Te$_3$/EuS/Se sample.

The negligible induced magnetic moment on Bi, Sb, Te, and Se atoms
of (Bi,Sb)$_2$(Se,Te)$_3$ in contact with EuS is a surprising
finding. Recent observations suggest the presence of magnetism
within the TI layer of Bi$_2$Se$_3$/EuS by magnetotransport
measurements \cite{WeiPRL2013, YangPRB2018},  polarized neutron
reflectometry \cite{KatmisNat2016, LiPRB2017}, magnetic
second-harmonic generation \cite{LeeNatComms2016} and low-energy
muon spin rotation \cite{Krieger_PRB2019}. Katmis \textit{et
al.}~\cite{KatmisNat2016} reported values of magnetization in the TI
for a Bi$_2$Se$_3$(20 QL)$/$EuS(5 nm) sample between 250 and 300
emu/cm$^3$, which correspond to a magnetic moment between 4.4 and
5.3 $\mu_\mathrm{B}$ per QL in a unit cell of Bi$_2$Se$_3$. Assuming
that Bi and Se atoms become equally magnetic, and that magnetism
develops over a unit cell closest to EuS, the magnetic moment
amounts to $\sim$0.88-1.06 $\mu_\mathrm{B}$ per (Bi or Se) atom,
which is about 2 to 3 orders of magnitude larger than those
estimated for our Bi$_{2}$Se$_{3}$/EuS samples from XMCD
measurements.

XMCD measurements have already been used to rule out reported
proximity effects using other (non-element selective) experimental
techniques. For example, the anisotropic magnetoresistance at
Y$_3$Fe$_5$O$_{12}$/Pt and ferrite/Pt interfaces was attributed to
Pt atoms polarized by proximity effects. However, XMCD measurements
at the Pt edges have shown no evidence of induced magnetism
\cite{GepragsAPL2012, ValvidaresPRB2016, ColletAPL2017}. Theoretical
studies using first-principles calculations show insignificant or no
magnetism in the TI for Bi$_2$Se$_3$/EuS with ideally sharp
interfaces \cite{EremeevJMMM2015, KimPRL2017, Meyerheim_2020}, which
is consistent with our XMCD results. Instead, Eu doping on
Bi$_2$Se$_3$ could result in local magnetic moments that are much
larger than those induced by the adjacent magnetic layer, possibly
explaining the experimental observations \cite{EremeevJMMM2015}. One
Eu atom diffusing into the TI per unit cell would account for an
observed magnetic moment of $\sim$1 $\mu_\mathrm{B}$/at in a QL
\cite{EremeevJMMM2015}. Intercalation of a single
EuS layer has been theoretically predicted to be energetically
favorable and leads to the observation of magnetism within the TI
and, perhaps, of proximity effects, by suppressing the formation of
trivial states \cite{EremeevNanolett2018}. However, our data does
not show evidence of such an intercalation. Other interfacial
effects, including charge transfer at the TI/MI interface and band
bending could explain the absence of induced magnetism in the TI and
the lack of observation of the QAHE
\cite{EremeevNanolett2018,MenshovPRB2013, EremeevPRB2013,
MenshovPRB2019, PetrovJETPLett2019}. Another possibility for the
absence of proximity effects would be the presence of a magnetic
dead layer at the TI/EuS interface. However, XAS and XMCD
measurements of ultrathin EuS layers ($\leq1$ nm) grown onto
Bi$_2$Se$_3$ reveal that EuS remains magnetic, which allowed us to
exclude such scenario in our TI/EuS samples \cite{SupplInfo}. This
conclusion is further supported by the increase of the in-plane
magnetic remanence for the TI/EuS structures, compared to the
isolated EuS layer, which can be ascribed to the large intrinsic
spin-orbit interaction of the TI at a sharp TI/EuS interface
\cite{KimPRL2017}.


In summary, we have performed element-selective magnetometry to
study proximity effects at the interface of TI/EuS bilayers. A
variety of TIs of the (Bi,Sb)$_2$(Se,Te)$_3$ family were
systematically investigated. The easy-magnetization axis of EuS remains in the plane of
the sample and the magnetic remanence increases as a result of its contact with
the TI. However, we found no evidence for an enhancement
of Eu magnetic moments in the EuS layer, or an increase in the
ferromagnetic ordering temperature. The magnetic signals on Bi, Sb, Te, and
Se atoms of the TIs in contact with EuS were found to be negligible (below the
detection limit of the XMCD technique) at temperatures and
conditions where the EuS is magnetic and could polarize these
non-magnetic atoms. Overall our XMCD measurements suggest that the
observations in TI/EuS interfaces by magnetotransport and optical
techniques are not due to proximity-induced magnetism but to a
different mechanism, such as magnetic doping by Eu diffusion into
the TI.

This research was supported by the European Union's Horizon 2020
FET-PROACTIVE project TOCHA under grant agreement 824140. The
authors acknowledge funding by the European Research Council under
grant agreement no.~306652 SPINBOUND, from the CERCA
Programme/Generalitat de Catalunya 2017 SGR 827 and by MINECO (under
contracts no.~MAT2016-75952-R, MAT2016- 78293-C6-2-R, PID2019-111773RB-I00 / AEI / 10.13039/501100011033, and Severo
Ochoa no.~SEV-2017-0706). A.~I.~F.\ acknowledges funding from
MINECO-Juan de la Cierva fellowship ref.~IJCI-2015-25514 and from
the European Union's Horizon 2020 People Programme (Marie
Sk\l{}odowska Curie Actions) H2020-MSCA-IF-2017 under REA grant
agreement No.~796925. F.~B.\ acknowledges funding from MINECO
Ram\'{o}n y Cajal Program under Contract No.~RYC-2015-18523. We
acknowledge beamtime on BOREAS (BL29) beamline at ALBA Synchrotron
under proposal 2017022019.


\providecommand{\noopsort}[1]{}\providecommand{\singleletter}[1]{#1}%

\end{document}